\documentclass[reprint,aps,prl,showpacs,amsmath,amssymb,floatfix,superscriptaddress]{revtex4-1}

\usepackage{times,graphicx,color,placeins}
\usepackage[colorlinks=true,urlcolor=blue,linkcolor=blue,
            citecolor=blue]{hyperref}

\newcommand{\bibtitle}[1]{\textit{#1},}
\DeclareMathOperator*{\argmin}{arg\,min}

\begin{document}

\title{Evidence from quantum Monte Carlo of large gap superfluidity
and BCS-BEC crossover in double electron-hole layers}

\author{Pablo L\'opez R\'ios}
  \email{pl275@cam.ac.uk}
  \affiliation{Max-Planck Institute for Solid State Research,
               Heisenbergstra{\ss}e 1, 70569 Stuttgart, Germany}
  \affiliation{Theory of Condensed Matter Group, Cavendish Laboratory,
               19 J. J. Thomson Avenue, Cambridge CB3 0HE, UK}

\author{Andrea Perali}
  \affiliation{School of Pharmacy, Physics Unit,
               University of Camerino, 62032 Camerino (MC), Italy}

\author{Richard J. Needs}
  \affiliation{Theory of Condensed Matter Group, Cavendish Laboratory,
               19 J. J. Thomson Avenue, Cambridge CB3 0HE, UK}

\author{David Neilson}
  \affiliation{School of Science and Technology, Physics Division,
               University of Camerino, 62032 Camerino (MC), Italy}
  \affiliation{Department of Physics, University of Antwerp,
               Groenenborgerlaan 171, 2020 Antwerp, Belgium}

\begin{abstract}
  We report quantum Monte Carlo evidence of the existence of large gap
  superfluidity in electron-hole double layers over wide density
  ranges.
  The superfluid parameters evolve from normal state to BEC with
  decreasing density, with the BCS state restricted to a tiny range of
  densities due to the strong screening of Coulomb interactions, which
  causes the gap to rapidly become large near the onset of
  superfluidity.
  The superfluid properties exhibit similarities to ultracold fermions
  and iron-based superconductors, suggesting an underlying universal
  behavior of BCS-BEC crossovers in pairing systems.
\end{abstract}


\maketitle


There are intense ongoing experimental efforts to observe
superfluidity in electron-hole double layer systems, including double
quantum wells in GaAs-AlGaAs heterostructures, double graphene
monolayers, double graphene bilayers, and hybrid graphene-GaAs
structures \cite{Sivan, Croxall, Seamons, Gorbachev, Mink}.
A recent very significant experimental advance by multiple groups has
been the fabrication of closely spaced electron-hole double graphene
bilayers with carrier densities tunable by metal gates
\cite{tutuc2016, dean2016, kim2017, dean2017}.
Insertion of a few layers of hexagonal boron nitride between the
bilayers creates an insulating barrier.
This can be as thin as $1$ nm while still blocking tunneling of
carriers between the bilayers, permitting the electrons and holes to
interact with a very strong Coulomb attraction.

Condensation of electron-hole pairs into a BEC superfluid state in
double graphene bilayers has been experimentally demonstrated in the
quantum Hall regime with a magnetic field \cite{kim2017, dean2017},
opening the way to the generation of quantum coherent macroscopic
states in spatially separated two-dimensional sheets.

Room temperature superfluidity had in fact been predicted earlier in
electron-hole double graphene monolayers \cite{Min}, but it was later
established that since a graphene monolayer remains blocked in the
weakly-coupled regime, strong screening always suppresses the
superfluidity \cite{Lozovik}.
Double electron-hole graphene bilayers were subsequently proposed to
overcome this problem \cite{Perali}, combining ideas for the
realization of high-$T_c$ superfluidity with the ability to move
across the BCS-BEC crossover by changing the carrier densities using
metal gates.
This ability to tune the system into the strong-coupling regime is key
to obtaining a superfluid in a solid-state device at experimentally
accessible conditions \cite{Perali}, offering a fascinating
alternative to ultracold fermionic atoms for studying superfluid
physics across the BCS-BEC crossover.

With superfluidity in cold atoms, quantum Monte Carlo (QMC) results
\cite{PeraliQMC2004, Bulgac2008, Morris2010} were from the outset
integrated closely with experiments and theory to understand and
control the phenomenon.
QMC simulations are extremely useful in any strongly correlated
condensed matter system where there is no small parameter that can be
used in perturbative expansions or controllable diagrammatic
approximations.
For double layer electron-hole systems, QMC simulations
\cite{depalo_2002, Maezono} are of key importance since the superfluid
phase arises from a complicated competition between the long-ranged
Coulomb interlayer attraction and intralayer repulsion.
Our work provides a comprehensive characterization of the superfluid
properties of electron-hole double layers at zero temperature, and
establishes the density range in which the condensate exists.
Our predictions that too high a density kills superfluidity and of the
existence of a BCS-BEC crossover regime at accessible intermediate
densities will be extremely useful in guiding experiments.
We use state-of-the-art QMC methods to determine the
dependence of BCS-like effective parameters on the equal carrier
densities, with the distance separating the layers fixed at the value
that maximizes exciton formation \cite{Maezono, Schindler_2008,
Lee_2009}.

The boundary between the crossover and BEC regimes is of great
interest because the pseudogap transforms into a real gap associated
with the molecular phase near this boundary \cite{Perali2011}.
The BCS-BEC crossover has been recently observed in the shallow Fermi
surface pockets of iron-based superconductors \cite{Kasahara, Hosono}.

The zero temperature BCS-BEC crossover \cite{Eagles, Leggett} can be
traced by following the evolution of the condensate fraction $c$, the
average radius of the superfluid pairs $r_{\rm ex}$ \cite{Salasnich},
or the superfluid gap $\Delta$.
Here we use explicit many-body wave-function-based methods to track
the evolution of the superfluid with carrier density in symmetric
electron-hole layers.
We directly obtain total energies, condensate fractions, and pair
radii, and we extract superfluid gaps and chemical potentials from
total energies using BCS-like relations with effective parameters.
We employ methods similar to those applied in Ref.\@
\onlinecite{Matveeva} for coupled layers of dipolar fermions.

The condensate fraction $c$ measures the fraction of carriers in the
condensate \cite{yang_1962, Regal2004, Perali2005, Manini}.
In the BCS regime $c<0.2$, with only a small fraction of the carriers
near the Fermi surface forming the condensate, while in the BEC regime
$c>0.8$, so almost all the carriers form local molecular bosonic
pairs, and condense.
Using the pair radius $r_{\rm ex}$, the BCS regime is characterized by
$k_{\rm F} r_{\rm ex}\gg 1$, where $k_{\rm F}$ is the Fermi wave
vector, while in the BEC regime the compact pairs correspond to
$k_{\rm F}r_{\rm ex}<1$.
The parameter $k_{\rm F}r_{\rm ex}$ was the first to be studied in
investigations of the BCS-BEC crossover in the high-$T_c$ cuprate
superconductors \cite{Pistolesi}.
$r_{\rm ex}$ determines the correlation length of the pairs, which
enters the expressions for the properties of the vortex state of the
superfluid and all other quantities affected by the spatial structure
of the superfluid wave-function, such as Josephson and Andreev
reflection effects \cite{Spuntarelli_2010}.

In contrast to $c$ and $r_{\rm ex}$, the superfluid gap is
experimentally straightforward to measure using angle-resolved
photoemission spectroscopy (ARPES), scanning tunnelling microscopy, or
measurements of the specific heat.
Knowledge of the evolution of the superfluid gap with the external
parameters is of fundamental relevance in designing experiments to
detect and characterize electron-hole superfluidity.
It is possible to link the entry into the BCS-BEC crossover regime as
determined by $c$ and $k_{\rm F}r_{\rm ex}$ to $\Delta/E_{\rm F}\sim
1$ at zero temperature \cite{Guidini}.
In our calculations we also monitor the evolution of the
pseudo-Luttinger wave vector $k_{\rm min}$ at which the $k$-dependent
excitation energy passes through its minimum.
$k_{\rm min}$ can be traced by ARPES, because it directly affects the
shape of the remnant Fermi surface in the broken symmetry phase at
zero temperature.

With our results for $\Delta$, $\mu$, $c$, and $r_{\rm ex}$ as
functions of density, we are then in a position to follow the
evolution of the system through the weak-coupling regime, the
superfluid BCS-BEC crossover regime, and the BEC regime, enabling
comparisons with predictions from various microscopic theories.


In our calculations we use excitonic Hartree units, $\hbar = |e| =
m_e^* = \kappa 4\pi\epsilon_0 = 1$, where $m_e^*$ is the effective
electron mass and $\kappa$ is the relative permittivity of the system,
and we obtain energies in units of ${\rm Ha}^* = (m_e^*/m_e)
\kappa^{-2}\,{\rm Ha}$ and distances in units of $a_0^* = \kappa
(m_e/m_e^*) \, a_0$.
For reference, the relative permittivity for bilayer graphene (BLG)
encapsulated in few-layer hexagonal boron nitride (hBN) is $\kappa=2$
\cite{Kumar2016}, and the effective mass is $m_e^* = 0.04 m_e$
\cite{Zou}.

We simulate a finite version of the paramagnetic, equal-mass
electron-hole double layer system with parabolic single-particle
energy dispersion using square simulation cells of area $A$ subject to
periodic boundary conditions, with $N$ particles in each layer.
The in-layer particle density is defined via the density parameter
$r_{\rm s}=\sqrt{A/(\pi N)}$.
We run calculations for systems containing $N=58$ electron-hole pairs;
tests with systems of $N=114$ electron-hole pairs show that
finite-size errors are small in our results \cite{supplemental}.
We use a fixed interlayer separation of $d/a_0^*=0.4$, slightly
greater than the largest $d$ at which biexciton formation is favorable
\cite{Maezono, Schindler_2008, Lee_2009}, and vary the density between
$r_{\rm s}/a_0^* = 1.75$ and $15$.

At all densities considered we evaluate total energies, condensate
fractions, and pair-correlation functions (PCFs) using the variational
quantum Monte Carlo (VMC) method \cite{mcmillan_1965,
foulkes_rmp_2001}.
Wave function parameters can be optimized within VMC
\cite{toulouse_emin_2007, umrigar_emin_2007}, and the accuracy of VMC
expectation values depends on the quality of the resulting wave
function.
The more computationally costly diffusion quantum Monte Carlo (DMC)
method \cite{ceperley_1980, foulkes_rmp_2001} employs stochastic
projection to extract the lowest-energy state compatible with the
nodal surface of a VMC-optimized trial wave function.
Once the time-step and population-control biases are eliminated
\cite{Vrbik_1986, Lee_2011, Zen_2016}, the accuracy of the DMC method
depends only on the accuracy of the nodes of the trial wave function.
We have performed DMC calculations at selected densities
representative of the weak-coupling, crossover, and strong-coupling
regimes.
Thus, our DMC calculations serve as quantitative corrections to our
VMC results throughout the density range considered.

We use trial wave functions of the form $\Psi = e^J D_{e\uparrow
h\downarrow} D_{e\downarrow h\uparrow}$, where $e^J$ is a Jastrow
correlation factor \cite{ndd_jastrow, gjastrow}, imposing the Kato
cusp conditions \cite{kato_1957}, and $D_{e\uparrow h\downarrow}$ and
$D_{e\downarrow h\uparrow}$ are pairing determinants
\cite{depalo_2002, Maezono}.
In systems with an additional (up-spin) electron, we complete the
corresponding determinant with a plane-wave orbital of wave vector
$\bf k$.
Details of our trial wave functions are given in the Supplemental
Material \cite{supplemental}.
We use the \textsc{casino} code for our calculations
\cite{casino_reference}.

The main properties of the BCS-BEC crossover at low temperature are
captured by BCS theory, as demonstrated with ultracold fermions.
When the gap equation is coupled to the density equation, the
excitation energy $\varepsilon(k)$ corresponding to the addition of an
electron of wave vector ${\bf k}$ follows the BCS dispersion relation
\cite{Matveeva},
\begin{equation}
  \label{eq:epsilon_BCS}
  \varepsilon(k) = \sqrt{ \left( {k^2}/{2m^*} - \mu \right)^2
                          + \Delta^2} \;,
\end{equation}
where $\mu$, $m^*$, and $\Delta$ are the chemical potential, effective
mass, and superfluid gap of the electron quasiparticle, respectively.
$\varepsilon(k)$ therefore contains the parameters that characterize
the superfluid state.
This excitation energy can be obtained from \textit{ab initio}
total energies as
\begin{equation}
  \label{eq:epsilon_QMC}
  \varepsilon(k) = E_A(N+1/2;k) - E_A(N) - \mu_{\rm QMC}(N) \;,
\end{equation}
where $E_A(N)$ is the energy of a system of area $A$ containing $N$
electron-hole pairs, $E_A(N+1/2;k)$ is the energy of a system of area
$A$ containing $N$ electron-hole pairs and one additional electron of
associated wave vector $\bf k$, and $\mu_{\rm QMC}(N) \approx \frac 1
4 \left[ E_A(N+1) - E_A(N-1) \right]$ is the chemical potential of the
system.
Note that $\mu_{\rm QMC}$ differs from the mean-field $\mu$ due to
many-body effects.

\begin{figure}[htb!]
  \includegraphics[width=0.47\textwidth]{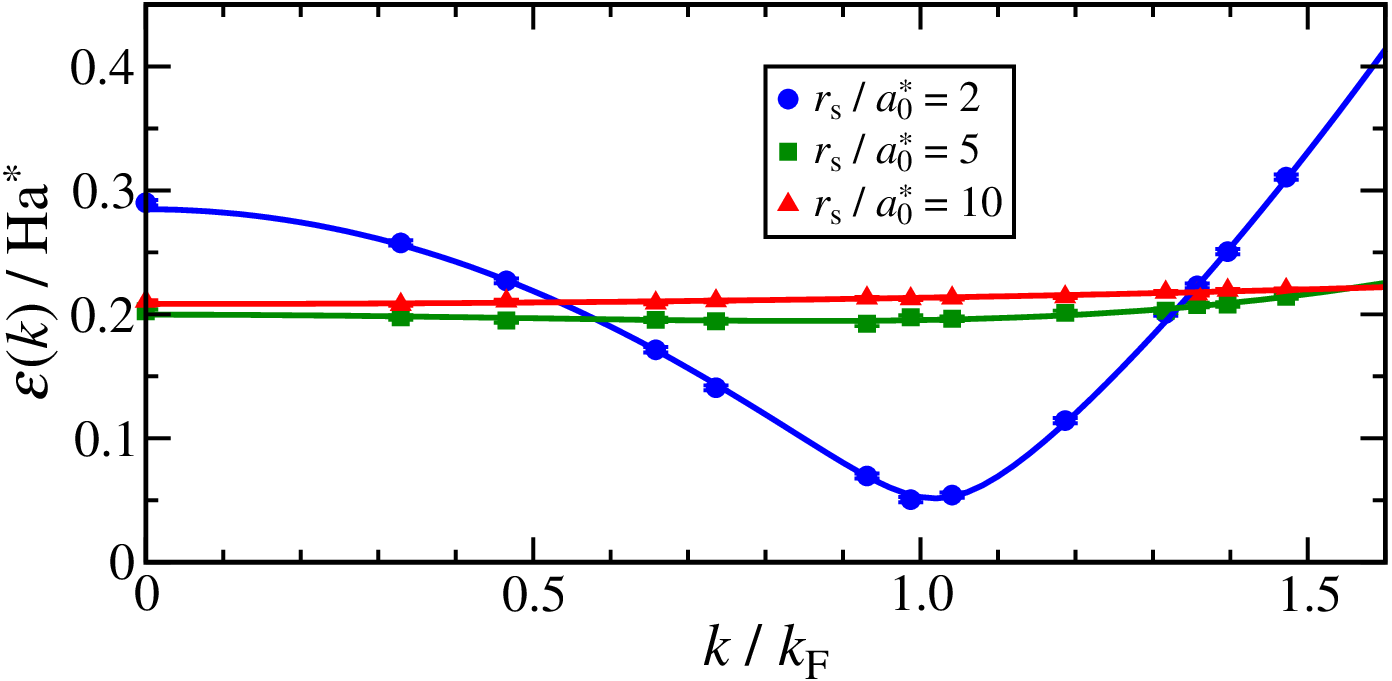}
  \caption{
    DMC estimates of $\varepsilon(k)$ as a function of the magnitude
    $k$ of the wave vector of the additional electron, at
    $r_{\rm s}/a_0^* = 2$, $5$, and $10$.
    The solid lines are fits of the DMC data to Eq.\@
    \ref{eq:epsilon_BCS}.
  }
  \label{epsilon_k}
\end{figure}

We simulate systems with $N$, $N+1$, and $N-1$ electron-hole
pairs, and systems with $N$ electron-hole pairs and an unpaired
electron at several wave vectors $\bf k$.
We then compute $\varepsilon(k)$ at each $\bf k$ using Eq.\@
\ref{eq:epsilon_QMC} and fit the resulting values to Eq.\@
\ref{eq:epsilon_BCS} with $m^*$, $\mu$, and $\Delta$ as fitting
parameters.
Following tests \cite{supplemental}, we use wave vectors such that
$0\leq k<k_{\rm cut}$; we set $k_{\rm cut}/k_{\rm F} = 1.5$
at high densities and use larger cut-off values at low densities.
Although the superfluid gap is expected to be $k$-dependent
\cite{Lozovik, Perali}, our calculations do not yield any significant
variation of $\Delta$ with $k$ in the ranges of $k$ we have considered
\cite{supplemental}.
Figure \ref{epsilon_k} shows a plot of the DMC values of
$\varepsilon(k)$ for densities of $r_{\rm s}/a_0^*=2$, $5$, and $10$,
along with the resulting fits to Eq.\@ \ref{eq:epsilon_BCS}.
The fits follow the DMC data remarkably well, indicating that the
BCS dispersion relation provides a robust description of the
\textit{ab initio} results.

\begin{figure}[htb!]
  \includegraphics[clip,width=0.47\textwidth]{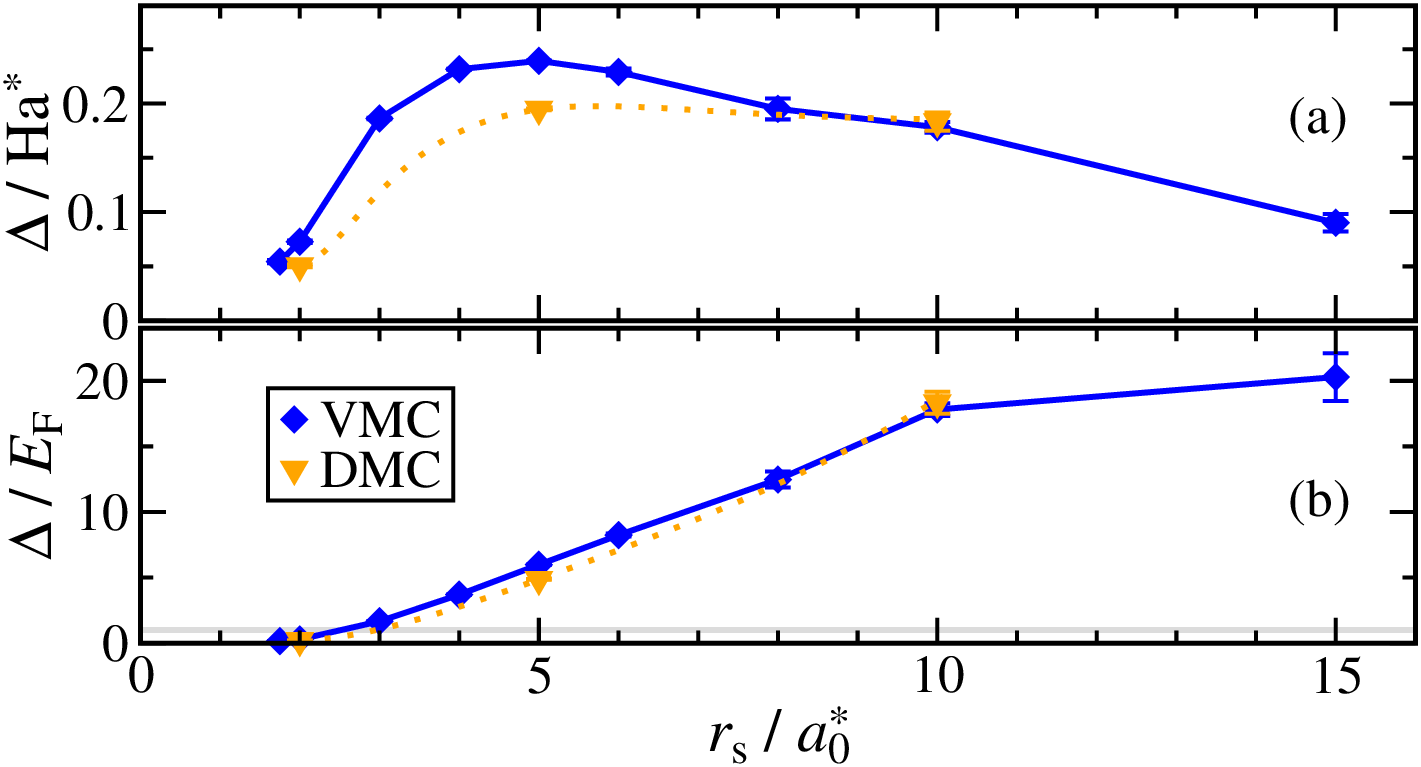}
  \caption{
    Superfluid gap $\Delta$, obtained by fitting the VMC and DMC data
    to Eq.\@ \ref{eq:epsilon_BCS}, as a function of $r_{\rm s}$, (a)
    in excitonic Hartree units, and (b) relative to $E_{\rm F}$.
    The dotted lines interpolating the DMC results are intended as a
    guide to the eye.
  }
  \label{fit_delta}
\end{figure}

In Fig.\@ \ref{fit_delta} the zero-temperature superfluid gap is
reported as a function of $r_{\rm s}$, both in excitonic Hartree units
and relative to $E_{\rm F}$.
At high densities there is no superfluidity because the electron-hole
pairing interaction is strongly screened \cite{Perali}.
Near the onset density, the electron-hole condensate is already close
to the BCS-BEC crossover boundary.
As $r_{\rm s}/a_0^*$ increases above $2$ there is a very steep
increase in the gap, which exceeds $\Delta / E_{\rm F} \sim 1$ by
$r_{\rm s}/a_0^* \sim 3$.
$\Delta / E_{\rm F} > 1$ signals entry into the BCS-BEC crossover
regime, so this occurs practically immediately after the onset of
superfluidity.
The steep rise in $\Delta$ is associated with strong screening at high
densities \cite{Perali}.
Consequently the weakly coupled BCS superfluidity regime, for which
$\Delta/E_{\rm F} \ll 1$, exists at most in a tiny range of densities.

\begin{figure}[htb!]
  \includegraphics[clip,width=0.47\textwidth]{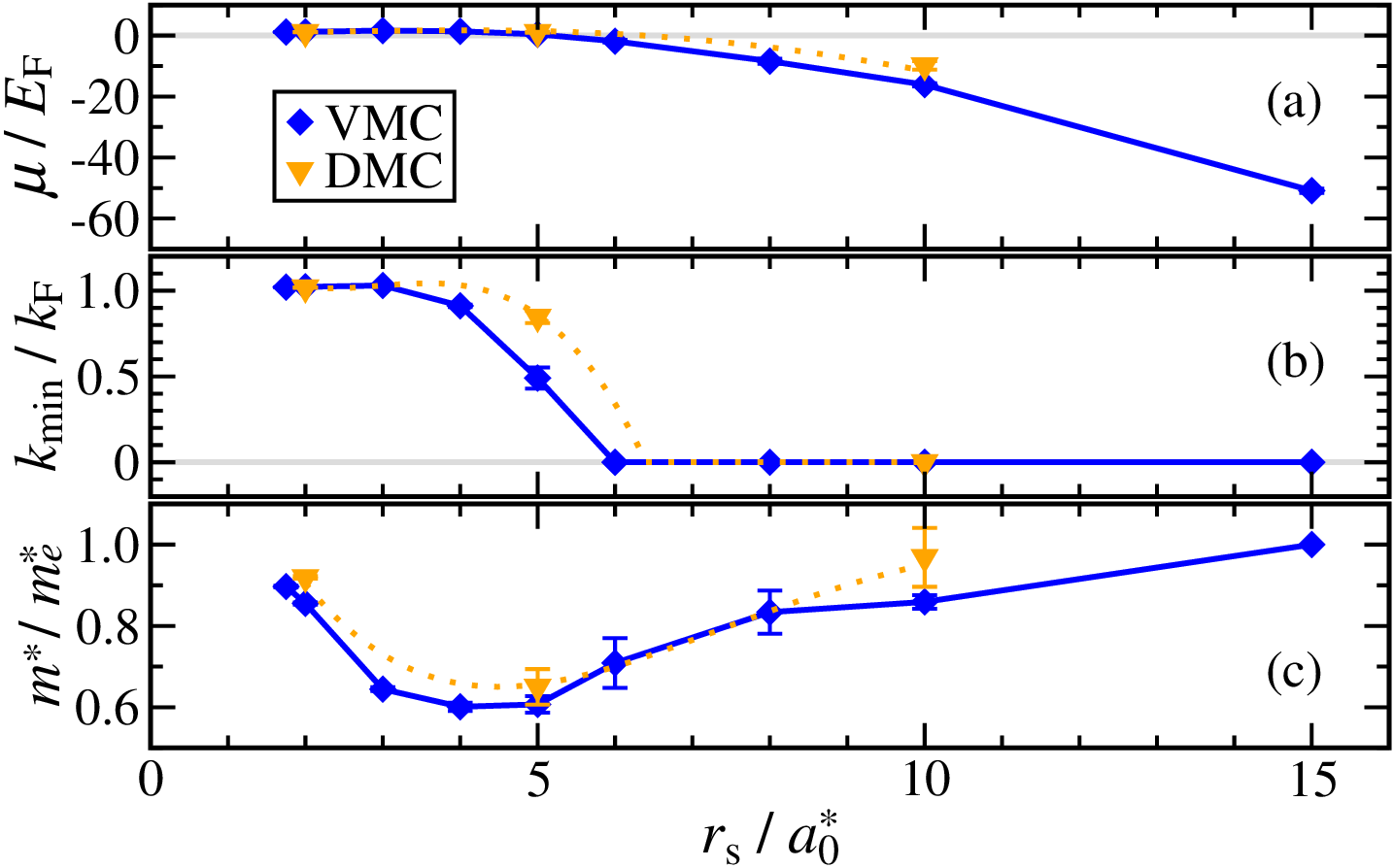}
  \caption{
    (a) $\mu$, (b) $k_{\rm min}$, and (c) $m^*$, obtained by fitting
    the VMC and DMC data to Eq.\@ \ref{eq:epsilon_BCS}, as functions
    of $r_{\rm s}$.
    The dotted lines interpolating the DMC results are intended as a
    guide to the eye.
  }
  \label{fit_params}
\end{figure}

Figure \ref{fit_params}\,(a) shows $\mu$ as a function of $r_{\rm s}$.
$\mu$ becomes negative by $r_{\rm s}/a_0^*\sim 6$, which signals entry
into the BEC regime.
In Figs.\@ \ref{fit_params}\,(b) and \ref{fit_params}\,(c) we plot
the location of the minimum of $\varepsilon(k)$,
$k_{\rm min} = \argmin_k \varepsilon(k)$, and $m^*$ as functions of
$r_{\rm s}$.
The value of $k_{\rm min}/k_{\rm F}$ tracks the collapse of the Fermi
surface, going from unity in the weak-coupling regime to zero in the
BEC regime.
The DMC results suggest that the Fermi surface fully collapses at a
somewhat lower density than predicted by VMC.

In Fig.\@ \ref{fit_params}\,(c) the quasiparticle mass $m^*$ has a
minimum of less than the effective electron mass $m_e^*$ near
where the superfluid gap is maximal, $r_{\rm s}/a_0^* \sim 5$.
This is indicative of the interplay between the intralayer repulsion
and the interlayer attraction, leading to quasiparticles in the
superfluid state with masses $m^*<m_e^*$ for intermediate $r_{\rm s}$.
This behavior of $m^*$ differs from theoretical and experimental
findings in ultracold fermions in two dimensions, where the
interaction is purely attractive.
There, $m^*\agt m_e^*$ always, and it varies monotonically with
$r_{\rm s}$.
The quasiparticle mass of the two-dimensional electron gas is
extensively discussed in Ref.\@ \onlinecite{drummond2009}.
Experimental measurements indicate a regime of small $r_{\rm s}/a_0^*
< 3$ in which the quasiparticle mass is smaller than the effective
electron mass, see Fig.\@ 4 of Ref.\@ \onlinecite{drummond2009}.
Thus, competition between intralayer repulsion and interlayer
attraction in the electron-hole double layer can lead to small
$m^*<m_e^*$, as we find.

\begin{figure}[htb!]
  \includegraphics[width=0.47\textwidth]{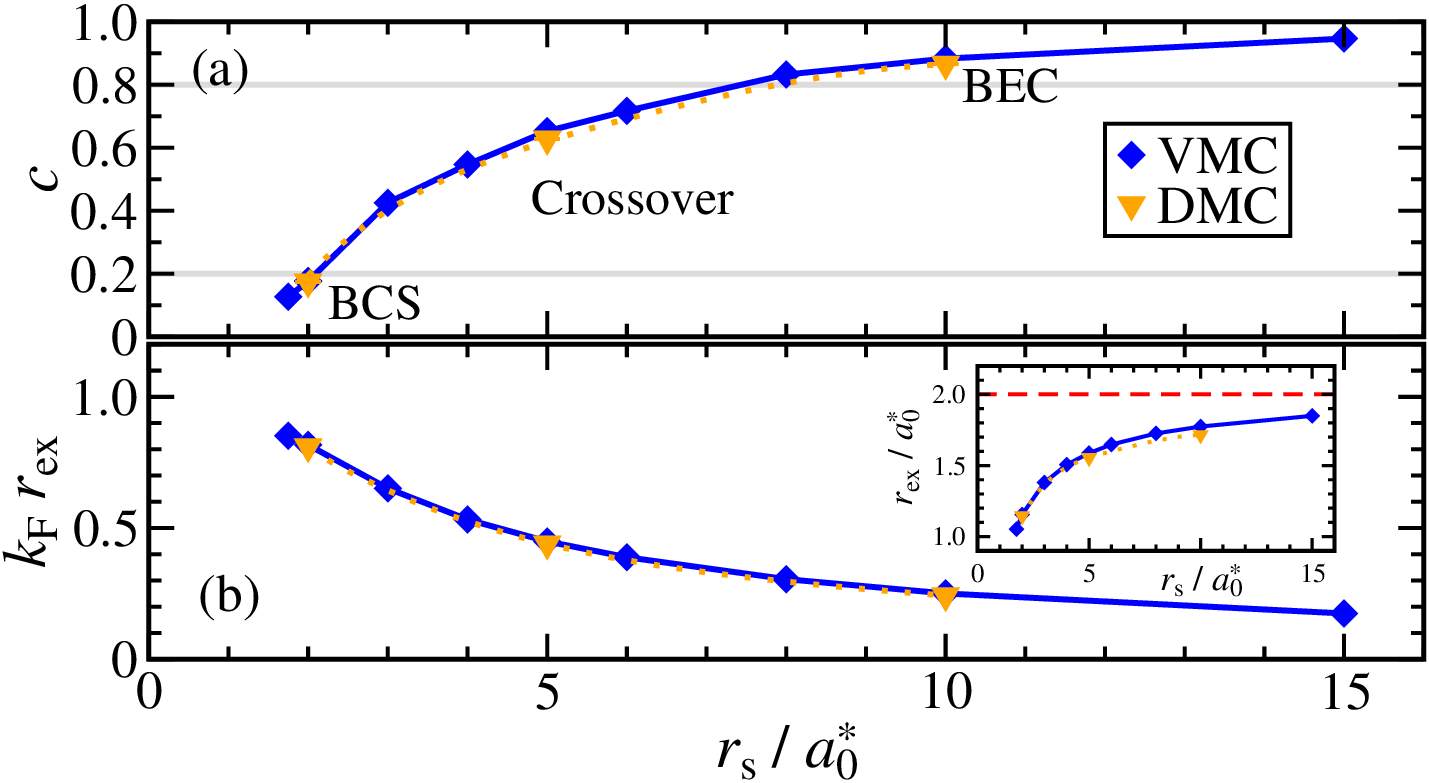}
  \caption{
    VMC and DMC values of (a) $c$ and (b) $r_{\rm ex}$ as functions of
    $r_{\rm s}$.
    The inset in the bottom panel shows $r_{\rm ex}$ in excitonic
    Hartree units.
    The dotted lines interpolating the DMC results are intended as a
    guide to the eye.
  }
  \label{expvals}
\end{figure}

The boundaries between the BCS, BCS-BEC crossover, and BEC regimes can
be determined from the condensate fraction $c$, which is defined as
\begin{equation}
  \label{eq:cfrac}
  c = (A^2/N)\lim_{r\to\infty} \rho_{eh}^{(2)}(r) \;,
\end{equation}
where $\rho_{eh}^{(2)}(r)$ is the translational-rotational average of
the two-body density matrix for electron-hole pairs
\cite{supplemental}.
We have evaluated $c$ using the estimator of Ref.\@
\onlinecite{Astrakharchik_2005} which removes one-body contributions
to ease extrapolation to the $r\to\infty$ limit.
The results for $c$ shown in Fig.\@ \ref{expvals}\,(a) are consistent
with the conclusions drawn from the behavior of $\mu$.
The condensate fraction is negligible for $r_{\rm s}/a_0^*\lesssim
1.5$ \cite{Maezono}.
As $r_{\rm s}/a_0^*$ increases, $c$ grows rapidly to $\sim
0.2$ by $r_{\rm s}/a_0^* = 2$, signaling entry into the BCS-BEC
crossover regime.
As $r_{\rm s}/a_0^*$ is further increased, $c$ increases substantially
and by $r_{\rm s}/a_0^*= 8$ it exceeds $c=0.8$, thus entering the BEC
regime.

We also compute the translational-rotational average of the
electron-hole PCF $g_{eh}(r)$ \cite{supplemental}, which allows us to
evaluate the exciton radius $r_{\rm ex}$ as
\begin{equation}
r_{\rm ex}^2 = \int_0^{r_1}
               r^2 g_{eh}(r) r_{\rm s}^{-2} \, 2\pi r \, {\rm d}r \;,
\end{equation}
where $r_1$ is the radius of a circle centred on a hole containing on
average one electron.
Figure \ref{expvals}\,(b) shows the pair radius $r_{\rm ex}$ for the
condensate.
As $r_{\rm s}$ increases, $r_{\rm ex}$ converges to the
isolated-exciton limit, which for $d/a_0^*=0.4$ is $r_{\rm ex}/a_0^*
= 2.002$.
The values of $k_{\rm F}r_{\rm ex}$ are always less than unity,
indicating pair sizes of the order of, or smaller than, the
interparticle distance, confirming that the superfluidity is always in
the strongly coupled crossover or BEC regimes.


In contrast with the short-range interactions typical of ultracold
fermions and superconductivity in general, electron-hole superfluidity
should be affected by screening because of the long-range nature of
the Coulomb pairing attraction.
The nature and effectiveness of this screening has been a source of
controversy in the past, with predictions from mean-field calculations
ranging from negligible screening resulting in room-temperature
superfluidity \cite{Min, Bistritzer, Lozovik, Sodemann}, to full
screening by the normal state which would essentially completely
suppress superfluidity \cite{Kharitonov}.
Reference \onlinecite{Neilson2014} compared the dramatically different
mean-field predictions for the density dependence of the condensate
fraction with the QMC values of the condensate fraction for the same
system \cite{Maezono}.
The conclusion was that the best mean-field approximation for
screening was self-consistent screening by the superfluid state
introduced by Lozovik \cite{Lozovik} and applied in Ref.\
\onlinecite{Perali} to double bilayer graphene.
Using a similar argument, Ref.\ \onlinecite{Neilson2014} concluded
that the vertex corrections and intralayer correlations in the
superfluid state make relatively small contributions.
Our present results for the superfluid gap, Fig.\
\ref{fit_delta}\,(a), are in good agreement with Fig.\ 3 of Ref.\
\onlinecite{Neilson2014}.
Our results are thus consistent with and further validate the
conclusion of Ref.\ \onlinecite{Neilson2014}.

The trends of our QMC results are consistent with the complete
suppression of superfluidity at high densities predicted in Ref.\@
\cite{Perali}.
After the onset of superfluidity, $r_{\rm s}/a_0^* = 1.5$, the system
is in the weak-coupled BCS regime, but the condensate fraction rises
rapidly and by $r_{\rm s}/a_0^* = 2$ it reaches $c=0.2$.
By $r_{\rm s}/a_0^* = 2$ the superfluid gap has reached $\sim E_{\rm
F}$.
Thus the BCS regime, for which $\Delta/ E_{\rm F}\ll 1$, is restricted
to the very small density range $1.5 < r_{\rm s}/a_0^* < 2$.
This confirms the effects of the highly non-trivial competition
between Coulomb screening, which tends to suppress electron-hole
pairing induced by the interlayer Coulomb attraction, and the opening
of a large superfluid gap, which suppresses the particle-hole
processes near the Fermi surface responsible for screening, thus
severely weakening the screening.

When $r_{\rm s}$ is further increased, the superfluid gap first
increases and then reaches a flat maximum around $r_{\rm s}/a_0^* \sim
5$ -- $6$ with a very large value $\Delta/{\rm Ha}^* \sim 0.2$.
The large gap indicates that the superfluidity is robust with a high
transition temperature.
The chemical potential is still positive at this density, $\mu/E_{\rm
F} \sim 0.5$, the condensate fraction is $\sim 0.7$, and $k_{\rm
min}/k_{\rm F} \sim 0.5$.
Thus the system retains its fermionic properties with a Fermi surface
intact but smeared out by the large gap, $\Delta/E_{\rm F} \sim 7$.

The BEC superfluid regime is reached at larger $r_{\rm s}/a_0^*\sim
8$, where the condensate fraction acquires values $c>0.8$.
In this regime $\mu/E_{\rm F}$ is large and negative, the ratio
$\Delta/E_{\rm F}>10$ is very large, the Fermi surface has completely
collapsed, and the average pair size approaches the radius of an
isolated exciton.
The electron-hole superfluid can then be regarded as an ensemble of
well-formed electron-hole dipoles, which are indirect excitons.
The excitons will behave as a two-dimensional bosonic gas with a
repulsive interaction, with a Kosterlitz-Thouless transition
\cite{kosterlitz_1973} governing the critical temperature for
superfluidity.
Thus in the BEC regime the critical temperature should diminish with
decreasing density.

We can thus conclude that the optimal density for experimental
realization of the electron-hole condensate is around $r_{\rm s}/a_0^*
\sim 5$, which is deep inside the BCS-BEC crossover regime with large
values of $\Delta/E_{\rm F}$.
For BLG encapsulated in hBN, $r_{\rm s}/a_0^* \sim 5$ corresponds to a
density of $2\times 10^{11}$ cm$^{-2}$, the maximum gap corresponds
to $\Delta \sim 54~{\rm meV} \sim 630~{\rm K}$, and the interlayer
distance $d/a_0^* = 0.4$ corresponds to $1$ nm.

Our QMC results are consistent with a universal behavior of materials
in the BCS, BCS-BEC, and BEC regimes not depending on the details of
the microscopic interactions, and thus they point to a very general
physics.
The ground state properties and their evolution with coupling strength
appear to be universal for (i) long-range Coulomb interactions, (ii)
contact interactions in fermions \cite{Perali2011}, and (iii) spin
fluctuations and phonons in iron-based superconductors
\cite{Kasahara, Hosono, Guidini}.

Our results confirm that, unlike for fermionic superfluids with
short-range pairing interactions, the BCS regime in Coulomb systems
with their long-range interactions and screening is restricted to a
very small range of densities.
This is due to competition between screening and the
superfluid gap \cite{Lozovik, Perali}, with strong screening
suppressing the small-gap BCS regime in Coulomb systems.
At high densities, the onset of superfluidity is delayed by screening,
so that when the onset density is eventually reached, the pairs are
relatively compact, and the superfluid gap, which rapidly becomes
large both in absolute value and relative to $E_{\rm F}$, will
strongly suppress screening.
Thus the system almost immediately enters the strong-coupling BCS-BEC
crossover regime.
For this reason the superfluidity is likely to be robust against
potential detrimental effects like disorder, density imbalance, and
low dimensional fluctuations, and we expect the largest gaps and
highest critical temperatures not to be far from the onset density.

\begin{acknowledgments}
  The authors thank G. Baym, M. Bonitz, and G. Senatore for useful
  discussions.
  A.P. and D.N. acknowledge financial support from University of
  Camerino FAR project CESEMN and from the Italian MIUR through the
  PRIN 2015 program under contract no.\@ 2015C5SEJJ001.
  R.J.N. acknowledges financial support from the Engineering and
  Physical Sciences Research Council, U.K., under grant no.\@
  EP/P034616/1.
  P.L.R. acknowledges financial support from the Max-Planck Society.
  Supporting research data may be freely accessed at
  \href{https://doi.org/10.17863/CAM.18830}
  {https://doi.org/10.17863/CAM.18830}, in compliance with the
  applicable Open Access policies.
  Computational resources have been provided by the High Performance
  Computing Service of the University of Cambridge and by the
  Max-Planck Institute for Solid State Research.
\end{acknowledgments}


\setcounter{table}{0}
\renewcommand{\thetable}{S\arabic{table}}
\setcounter{figure}{0}
\renewcommand{\thefigure}{S\arabic{figure}}
\setcounter{section}{0}
\renewcommand{\thesection}{S\arabic{section}}
\setcounter{equation}{0}
\renewcommand{\theequation}{S\arabic{equation}}
\renewcommand{\bibnumfmt}[1]{[S#1]}
\renewcommand{\citenumfont}[1]{S#1}

\pagebreak
\onecolumngrid
\begin{center}
\textbf{\large Evidence from quantum Monte Carlo of large gap
superfluidity and BCS-BEC crossover in double electron-hole
layers: Supplemental Material}
\end{center}
\twocolumngrid

\section{Hamiltonian and trial wave function}

The Hamiltonian of the paramagnetic, equal-mass, infinite
electron-hole double layer is, in excitonic Hartree units ($\hbar =
|e| = m_e^* = \kappa 4\pi\epsilon_0=1$) \cite{suppl_Maezono},
\begin{eqnarray}
\nonumber
\hat H &=& -\frac 1 2 \left( \sum_{i} \nabla_{{\bf e}_i}^2 +
                             \sum_{i} \nabla_{{\bf h}_i}^2 \right)
           + \sum_{i<j} \frac{1}{\left|{\bf e}_i-{\bf e}_j\right|} \\
\label{Seq:hamiltonian}
       & & + \sum_{i<j} \frac{1}{\left|{\bf h}_i-{\bf h}_j\right|}
           - \sum_{i,j} \frac{1}{
               \sqrt{d^2+\left|{\bf e}_i-{\bf h}_j\right|^2}}
           \;.
\end{eqnarray}
${\bf e}_i$ and ${\bf h}_j$ are the in-plane position vectors of the
$i$th electron and the $j$th hole and $d$ is the distance between the
layers.
Note that, in terms of the in-layer density parameter, $r_{\rm s}$,
the magnitude of the Fermi wave vector is $k_{\rm F} = \sqrt{2}/r_{\rm
s}$ and the Fermi energy is $E_{\rm F} = 1/r_{\rm s}^2$.
We simulate systems consisting of a finite number of electrons and
holes subject to periodic boundary conditions, thus the Coulomb
interaction of Eq.\@ \ref{Seq:hamiltonian} must be replaced with the
two-dimensional version of Ewald summations \cite{suppl_Parry_1976}.

Each of the pairing determinants in our trial wave function is of the
form
\begin{equation}
  \label{Seq:d_ref}
  D_{e\uparrow h\downarrow} =
    \left|
    \begin{array}{ccc}
      \phi({\bf e}_1^\uparrow - {\bf h}_1^\downarrow) &
      \cdots                                          &
      \phi({\bf e}_N^\uparrow - {\bf h}_1^\downarrow) \\
      \vdots & \ddots & \vdots \\
      \phi({\bf e}_1^\uparrow - {\bf h}_N^\downarrow) &
      \cdots                                          &
      \phi({\bf e}_N^\uparrow - {\bf h}_N^\downarrow)
    \end{array}
    \right|
    \;,
\end{equation}
where ${\bf e}_i^\sigma$ and ${\bf h}_i^\sigma$ are the position
vectors of the $i$th $\sigma$-spin electron and hole, respectively,
and $\phi({\bf r})$ is a pairing orbital containing optimizable
parameters of the form used in Ref.\@ \onlinecite{suppl_Maezono},
\begin{eqnarray}
  \nonumber
  \phi({\bf r}) & = &
    \sum_{l=1}^{41} p_l \cos({\bf k}_l \cdot {\bf r})
    \\ \label{Seq:mixed_pairing} & + &
    \left(1-r/L\right)^3 \Theta\left(r-L\right)
    \sum_{m=0}^{8} c_m r^m \;,
\end{eqnarray}
where ${\bf k}_l$ is the $l$th shortest reciprocal lattice vector
[excluding one of each $({\bf k},-{\bf k})$ pair due to symmetry]
and $\{p_l\}$, $\{c_m\}$, and $L$ are optimizable parameters,
with $p_l=p_{l^\prime}$ when $\left|{\bf k}_l\right| =
\left|{\bf k}_{l^\prime}\right|$.
The pairing orbital is constrained to be cuspless at ${\bf r} = {\bf
0}$.
There are $22$ optimizable parameters in the pairing orbitals (one of
the coefficients is fixed due to normalization, and another is
determined by the cusplessness condition).
At $r_{\rm s}/a_0^*\ge 8$ we find that the cosine expansion
does not provide useful variational freedom to the wave function and
we set $p_l=0$, in which case the pairing orbitals effectively contain
8 optimizable parameters.

We also run calculations for systems with an additional (up-spin)
electron for which we replace the first pairing determinant with
\begin{equation}
  \label{Seq:d_imb}
  D_{e\uparrow h\downarrow} =
    \left|
    \begin{array}{ccc}
      \phi({\bf e}_1^\uparrow - {\bf h}_1^\downarrow) &
      \cdots                                          &
      \phi({\bf e}_{N+1}^\uparrow - {\bf h}_1^\downarrow) \\
      \vdots & \vdots & \vdots \\
      \phi({\bf e}_1^\uparrow - {\bf h}_N^\downarrow) &
      \cdots                                          &
      \phi({\bf e}_{N+1}^\uparrow - {\bf h}_N^\downarrow) \\[0.1cm]
      \cos\left({\bf k}\cdot{\bf e}_1^\uparrow\right) &
      \cdots &
      \cos\left({\bf k}\cdot {\bf e}_{N+1}^\uparrow \right)
    \end{array}
    \right|
    \;,
\end{equation}
where ${\bf k}$ is a wave vector commensurate with the reciprocal
lattice of the simulation cell associated with the additional
electron.

We use an isotropic two-body Jastrow factor \cite{suppl_gjastrow} of the
Drummond-Towler-Needs parametrization \cite{suppl_ndd_jastrow},
$\exp\left[J(\bf R)\right] = \exp\left[\sum_{i<j} u(r_{ij})\right]$,
with
\begin{equation}
  \label{Seq:jastrow_u}
  u(r) =
    \left(r-L_u\right)^3 \Theta\left(r-L_u\right)
    \sum_{m=0}^8 \alpha_m r^m \;,
\end{equation}
where $r_{ij}$ is the distance between particles $i$ and $j$, $\Theta$
is the Heaviside step function, and $\{\alpha_m\}$ and $L_u$ are
optimizable parameters.
Different parameter values are used for same-spin same-particle pairs,
opposite-spin same-particle pairs and electron-hole pairs, and the
two-dimensional Kato cusp conditions are applied for same-particle
pairs \cite{suppl_kato_1957}.
The electron-hole $u$ function is constrained to be cuspless, and we
do not use the ``quasi-cusp'' electron-hole term decribed in Ref.\@
\onlinecite{suppl_Maezono}.
There are $27$ optimizable parameters in the Jastrow factor (the
cusp/cusplessness conditions determine one of the coefficients for
each particle-pair type).

\section{Evaluation of expectation values}

\begin{figure}[htb!]
  \includegraphics[width=0.48\textwidth]{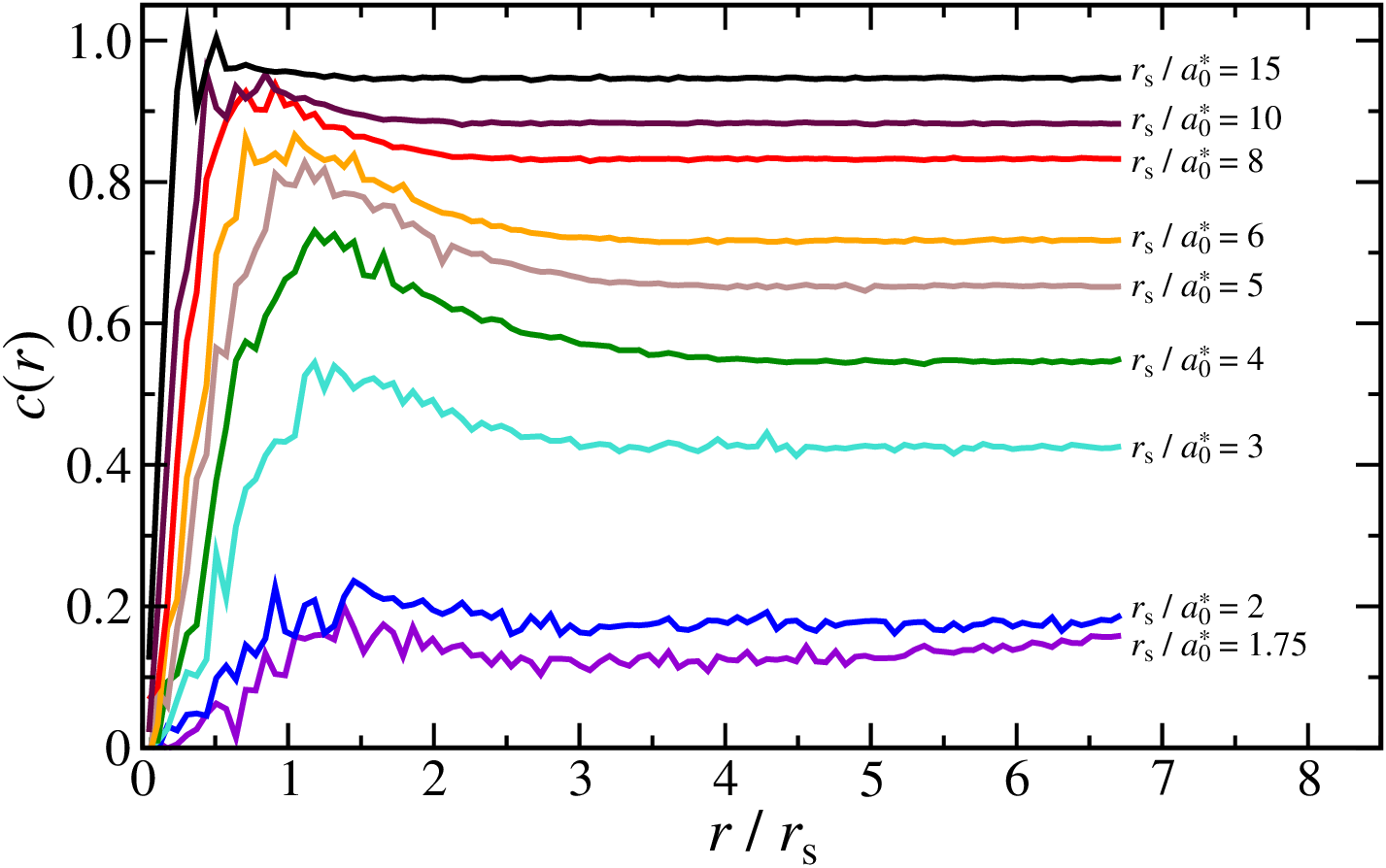}
  \caption{
    VMC estimate of the condensate fraction estimation function
    $c(r)$ as a function of $r/r_{\rm s}$ for the densities
    considered.
  }
  \label{Sfig:cfrac_function}
\end{figure}

The translational-rotational average of the one-body density matrix
for an electron (or hole) is
\begin{equation}
\label{Seq:1bdm_TR}
\rho_e^{(1)}(r) =
  \frac {N \int \left| \Psi({\bf R}) \right|^2
         \frac {\Psi({\bf e}+{\bf r}^\prime)}
               {\Psi({\bf e})}
         \delta(|{\bf r}^\prime|-r) \, {\rm d}{\bf R}
                                       {\rm d}{\bf r}^\prime }
        {A 2\pi r
         \int \left| \Psi({\bf R}) \right|^2 \, {\rm d}{\bf R} } \;.
\end{equation}
Similarly, the translational-rotational average of the two-body
density matrix for electron-hole pairs is
\begin{equation}
\label{Seq:2bdm_TR}
\rho_{eh}^{(2)}(r) =
  \frac {N^2 \int \left| \Psi({\bf R}) \right|^2
         \frac {\Psi({\bf e}+{\bf r}^\prime,
                     {\bf h}+{\bf r}^{\prime})}
               {\Psi({\bf e},
                     {\bf h})}
         \delta(|{\bf r}^\prime|-r) \, {\rm d}{\bf R}
                                       {\rm d}{\bf r}^\prime }
        {A^2 2\pi r
         \int \left| \Psi({\bf R}) \right|^2 \, {\rm d}{\bf R} } \;.
\end{equation}
The evaluation of $\rho_e^{(1)}(r)$ and $\rho_{eh}^{(2)}(r)$ in QMC is
performed by computing the corresponding wave function ratio at random
values of $r$ and accumulating the resulting values in bins.
We use these density matrices to compute the function
\cite{suppl_Astrakharchik_2005}
\begin{equation}
c(r) = \frac{A^2} N \left[
       \rho_{eh}^{(2)}(r) -
       \rho_e^{(1)}(r) \rho_h^{(1)}(r) \right] \;,
\end{equation}
from which the condensate fraction can be obtained as $c =
\lim_{r\to\infty} c(r)$.
In Fig.\@ \ref{Sfig:cfrac_function} we present VMC plots of $c(r)$ at
the densities considered, where the plateau at large $r$ that
determines $c$ can be seen.
The slight increase of the value of $c(r)$ at large $r$ for $r_{\rm s}
/ a_0^* = 1.75$ is typical of the weak-pairing regime.

\begin{figure}[htb!]
  \includegraphics[width=0.48\textwidth]{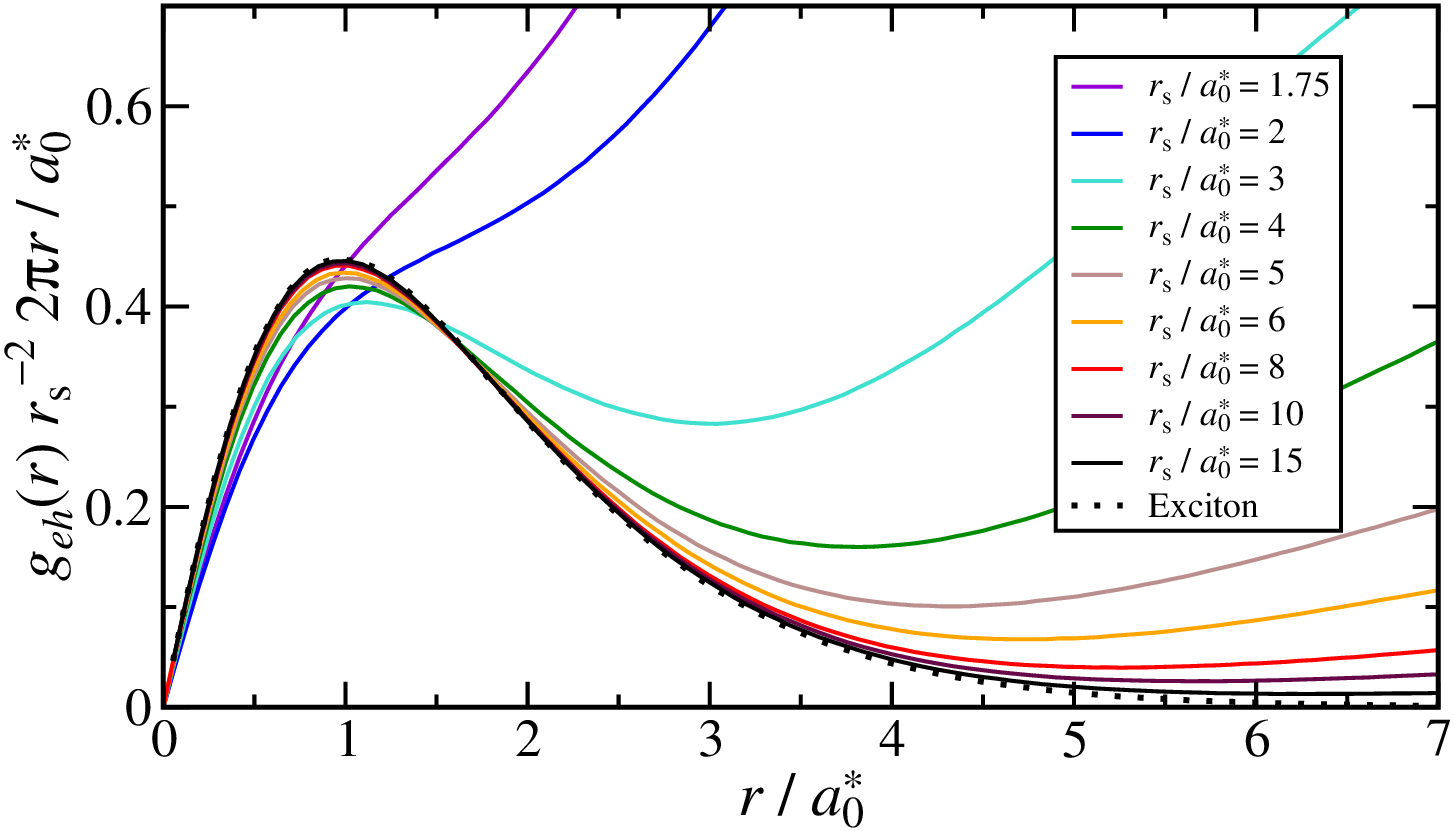}
  \caption{
    VMC estimate of the radial charge density $g_{eh}(r) r_{\rm
    s}^{-2} \, 2\pi r$ of electrons around a hole as a function of $r$
    for the densities considered, along with the radial charge
    density of an isolated exciton for comparison.
  }
  \label{Sfig:radial_charge_density}
\end{figure}

The translational-rotational average of the electron-hole
pair-correlation function (PCF) is
\begin{equation}
\label{Seq:pcf}
g_{eh}(r) =
  \frac{A
        \int |\Psi({\bf R})|^2 \delta\left({\bf e}-
             {\bf h}-{\bf r}^\prime\right)
             \delta\left(\left|{\bf r}^\prime\right|-r\right)
             \, {\rm d}{\bf R} {\rm d}{\bf r}^\prime}
       {2\pi r
        \int |\Psi({\bf R})|^2 \, {\rm d}{\bf R}} \;.
\end{equation}
We use the PCF to evaluate the average size of the exciton pair.
In Fig.\@ \ref{Sfig:radial_charge_density} we present VMC plots of
the radial charge density of electrons around a hole,
$g_{eh}(r) r_{\rm s}^{-2} \, 2\pi r$, at the densities considered,
as well as the radial charge density of an isolated exciton.

\section{Details of the diffusion Monte Carlo calculations}

We have performed DMC calculations at selected densities of
$r_{\rm s}/a_0^*=2$, $5$, and $10$.
Each DMC total energy is obtained from a DMC calculation consisting
of $M_1$ steps with a time step of $\tau_1$ and a target walker
population of $P_1$, and a second DMC calculation consisting of
$M_2=M_1/2$ steps with a time step of $\tau_2=4\tau_1$ and a target
walker population of $P_2=P_1/4$.
These ratios give the most efficient extrapolation of the results to
the zero-time-step, infinite-population limit for a fixed
computational cost \cite{suppl_Vrbik_1986, suppl_Lee_2011}.

The DMC time steps must be set according to the smallest length scale
of the system $\lambda$.
At high densities $\lambda$ is of the order of $r_{\rm s}$, while
at low densities $\lambda$ is of the order of the exciton radius.
Thus, we approximate $\lambda=\min(r_{\rm s},3 a_0^*)$ and set
$\tau_1 = 0.01 \lambda^2$.
We use $P_1=2048$ walkers and adjust $M_1$ to obtain the desired
statistical accuracy.

\begin{figure}[htb!]
  \includegraphics[width=0.48\textwidth]{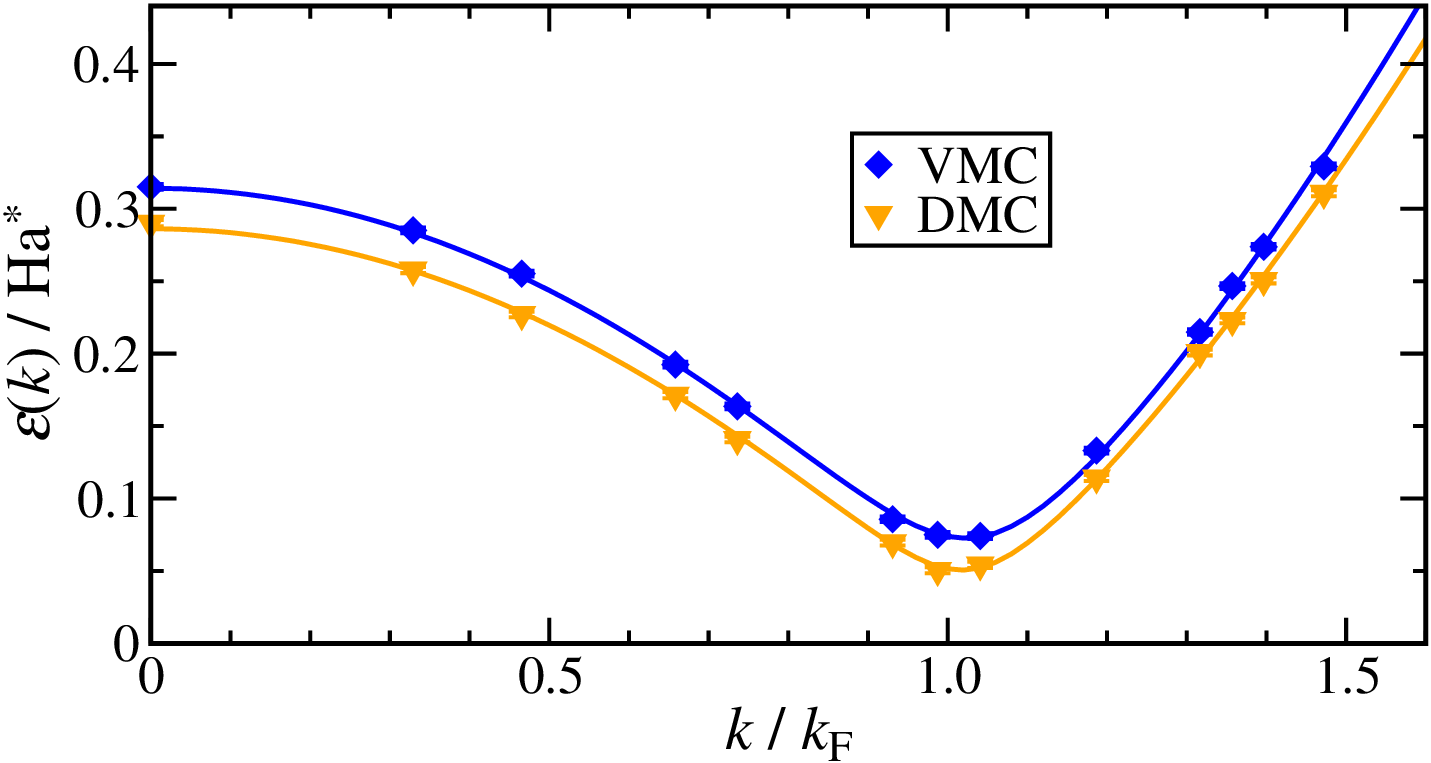}
  \caption{
    Excitation energies at $r_{\rm s}/a_0^*=2$ obtained
    from DMC (orange triangles) and from VMC (blue diamonds).
  }
  \label{Sfig:dmc_epsk_rs2}
\end{figure}

We compare the VMC and DMC excitation energies for $r_{\rm s}/a_0^*=2$
as a function of $k$ in Fig.\@ \ref{Sfig:dmc_epsk_rs2}.
The absolute differences between the VMC and DMC excitation energies
are of the order of $0.02~{\rm Ha}^*$, with the shape and basic
features of both curves being largely identical.

\section{Fitting procedure}

We evaluate the excitation energy using Eq.\@ 2 in the manuscript
at values of $k$ commensurate with the reciprocal lattice of the
simulation cell in the range $0\le k < k_{\rm cut}$, and fit the
results to the BCS dispersion relation, Eq.\@ 1 in the manuscript.
The standard errors in the fit parameters are determined by a
stochastic process where $\varepsilon(k)$ at each $k$ is replaced by
a random number drawn from a normal distribution centred at the mean
value of $\varepsilon(k)$ of variance its standard error, and the fit
is carried out.
We obtain the standard error in each fit parameter as the square root
of the variance of the values of the parameter in 10,000 realizations
of this process.

While the quality of a least-squares fit increases with the number of
data points available, we find that the fitting parameters depend
strongly on $k_{\rm cut}$ at small $r_{\rm s}$.
In particular, parameters $\mu$ and $m^*$ are intrinsic properties of
the electron quasiparticle and are expected to be $k$-independent, but
are found to vary significantly with $k_{\rm cut}$.
This dependence indicates that Eq.\@ 1 models the
\textit{ab initio} results inconsistently over different $k$ ranges,
and it is therefore critical to restrict $k_{\rm cut}$ so that the
values of $\mu$ and $m^*$ are constant over the whole range
$0\le k < k_{\rm cut}$.

We determine $k_{\rm cut}$ at each value of $r_{\rm s}$ by analyzing
subsets of VMC data in different $k$ ranges.
From the full set of $k$ values at which we have obtained
$\varepsilon(k)$, $k_1<k_2<\ldots<k_{n_k}$, we construct a subset of
$n$ consecutive data points from $k_i=k_{\rm L}$ to
$k_{i+n-1}=k_{\rm R}$, and perform a fit to Eq.\@
1.
We then compare the resulting fit parameters for different $k$ ranges,
and choose $k_{\rm cut}$ to be the largest $k_{\rm R}$ that yields
parameters $\mu$ and $m^*$ consistent with those from ranges with
smaller $k_{\rm R}$.

\begin{figure}[htb!]
  \includegraphics[width=0.48\textwidth]{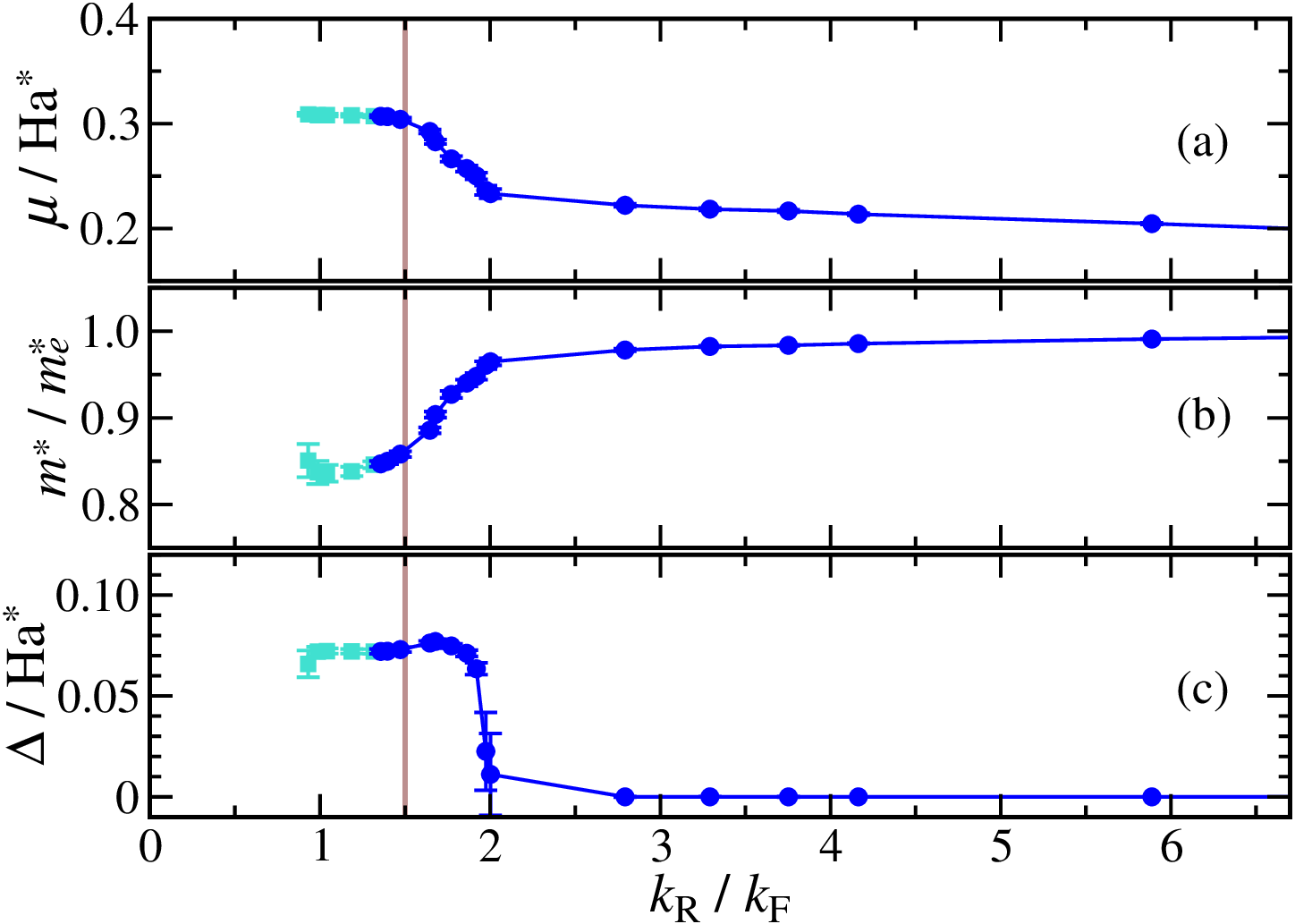}
  \caption{
    Values of the fit parameters (a) $\mu$, (b) $m^*$, and (c)
    $\Delta$ obtained by fitting subsets of $11$ consecutive VMC data
    points for $r_{\rm s}/a_0^*=2$ to Eq.\@ 1, as
    a function of the largest value of $k$ in each subset,
    $k_{\rm R}$.
    The five light-colored squares in each panel are the parameters
    obtained by fitting the first $6$, $7$, $8$, $9$, and $10$ data
    points, displayed here to show the stable behavior of the fit
    parameters in the low-$k$ region.
    The vertical line at $k_{\rm cut}/k_{\rm F}=1.5$ marks the point
    after which $\mu$ and $m^*$ start to deviate significantly from
    their low-$k$ values.
  }
  \label{Sfig:windowed_fit_rs2}
\end{figure}

In Fig.\@ \ref{Sfig:windowed_fit_rs2} we demonstrate this procedure
for $r_{\rm s}/a_0^*=2$ using $n=11$ points in each subset.
Also shown are fits using smaller windows containing the first $6$,
$7$, $8$, $9$, and $10$ data points to visualize the low-$k$ behavior
of the fit parameters.

For all densities in the range $1.75\le r_{\rm s}/a_0^*\le 6$ we find
that $\mu$ and $m^*$ remain constant up to $k_{\rm cut}/k_{\rm F} =
1.5$.
For $r_{\rm s}/a_0^*=8$ we set $k_{\rm cut}/k_{\rm F}=2$, and for
larger $r_{\rm s}$ we do not find any significant variation of the fit
parameters with $k_{\rm cut}$, and we use $n_k=40$ and $60$ data
points for $r_{\rm s}/a_0^*=10$ and $15$, respectively, corresponding
to $k_{\rm cut}/k_{\rm F} = 3.11$ and $3.97$.
We apply the values of $k_{\rm cut}$ obtained from this VMC assessment
to our DMC calculations, with the only difference that in our DMC
runs for $r_{\rm s}/a_0^* = 10$ we skip values of $k$ so as to
require $20$ data points instead of $40$, due to computational cost
considerations.

The superfluid gap $\Delta$ is predicted to be a function of $k$
peaked at $k_0$ of half-width $w$ and a $1/k$ tail at large $k$
\cite{suppl_Lozovik, suppl_Perali}.
The centre is expected to shift from $k_0 \sim k_{\rm F}$ in the BCS
regime to $k_0 = 0$ in the BEC regime, and the half-width is expected
to increase with $r_{\rm s}$.
However, due to the small $k$ region over which we are able to fit our
data to Eq.\@ 1, we do not detect any significant
variation of $\Delta$ with $k$, and we can only infer that its
half-width is of at least the order of $k_{\rm F}$ at high densities.
We note that the introduction of a $k$-dependent $\Delta$ of the
aforementioned properties does not account for the variation of
$\mu$ and $m^*$ with $k$ beyond $k_{\rm cut}$.

In Tables \ref{Stable:epsk_fit} and \ref{Stable:epsk_fit_dmc} we give
the parameters obtained using this fitting procedure with VMC and DMC
data, respectively, as plotted in the manuscript.
The value of the effective mass parameter appears to tend
asymptotically to $m_e^*$ as $r_{\rm s}$ increases.
At $r_{\rm s}/a_0^*=15$, since we obtain a VMC effective mass within
uncertainty of $m^*/m_e^*=1$, we perform a second fit with $m^*/m_e^*$
fixed at unity to improve the estimation of $\mu$ and $\Delta$, and we
report the parameters obtained from this fit as our final results for
this density.

\begin{table}[htb!]
  \centering
  \begin{tabular}{r@{.}lr@{.}lr@{.}lr@{.}lr@{.}lr@{.}l}
  \hline
  \hline
    \multicolumn{2}{c}{$r_{\rm s}/a_0^*$}
    & \multicolumn{2}{c}{$\Delta/{\rm Ha}^*$}
    & \multicolumn{2}{c}{$\mu/{\rm Ha}^*$}
    & \multicolumn{2}{c}{$m^*/m_e^*$}
    & \multicolumn{2}{c}{$c$}
    & \multicolumn{2}{c}{$r_{\rm ex}/a_0^*$} \\
  \hline
    $ 1$&$75$ & $0$&$0544(14)$ & $ 0$&$379(1) $ & $0$&$897(2) $
              & $0$&$1369(11)$ & $1$&$054$ \\
    $ 2$&$00$ & $0$&$0727(12)$ & $ 0$&$306(1) $ & $0$&$855(3) $
              & $0$&$1760(10)$ & $1$&$155$ \\
    $ 3$&$00$ & $0$&$1862(7) $ & $ 0$&$183(2) $ & $0$&$645(5) $
              & $0$&$4252(6) $ & $1$&$381$ \\
    $ 4$&$00$ & $0$&$2315(5) $ & $ 0$&$087(2) $ & $0$&$601(10)$
              & $0$&$5464(2) $ & $1$&$506$ \\
    $ 5$&$00$ & $0$&$2392(4) $ & $ 0$&$016(4) $ & $0$&$607(20)$
              & $0$&$6525(3) $ & $1$&$587$ \\
    $ 6$&$00$ & $0$&$2290(32)$ & $-0$&$051(12)$ & $0$&$709(61)$
              & $0$&$7171(2) $ & $1$&$648$ \\
    $ 8$&$00$ & $0$&$1950(96)$ & $-0$&$131(13)$ & $0$&$834(53)$
              & $0$&$8325(1) $ & $1$&$725$ \\
    $10$&$00$ & $0$&$1781(49)$ & $-0$&$161(5) $ & $0$&$859(17)$
              & $0$&$8828(1) $ & $1$&$774$ \\
    $15$&$00$ & $0$&$203(18) $ & $-0$&$220(8) $ & $0$&$972(27)$
              & $0$&$9467(1)$ & $1$&$849$ \\
    $15$&$00$ & $0$&$0902(81)$ & $-0$&$226(3) $ & $1$&$0      $
              {\it [fixed]} & \multicolumn{4}{c}{~}  \\
  \hline
  \hline
  \end{tabular}
  \caption{
     Superfluid parameters $\Delta$, $\mu$, and $m^*$ obtained by
     fitting VMC values of $\varepsilon(k)$ to Eq.\@
     1, along with VMC estimates of the
     condensate fraction $c$ and pair size $r_{\rm ex}$.
     \label{Stable:epsk_fit}}
\end{table}

\begin{table}[htb!]
  \centering
  \begin{tabular}{r@{.}lr@{.}lr@{.}lr@{.}lr@{.}lr@{.}l}
  \hline
  \hline
    \multicolumn{2}{c}{$r_{\rm s}/a_0^*$}
    & \multicolumn{2}{c}{$\Delta/{\rm Ha}^*$}
    & \multicolumn{2}{c}{$\mu/{\rm Ha}^*$}
    & \multicolumn{2}{c}{$m^*/m_e^*$}
    & \multicolumn{2}{c}{$c$}
    & \multicolumn{2}{c}{$r_{\rm ex}/a_0^*$} \\
  \hline
    $ 2$&$00$ & $0$&$0506(13)$  & $ 0$&$282(1) $ & $0$&$919(3) $
              & $0$&$17380(12)$ & $1$&$146$ \\
    $ 5$&$00$ & $0$&$1946(8) $  & $ 0$&$045(6) $ & $0$&$650(44)$
              & $0$&$62751(7) $ & $1$&$555$ \\
    $10$&$00$ & $0$&$1833(83)$  & $-0$&$098(14)$ & $0$&$969(73)$
              & $0$&$86542(1) $ & $1$&$715$ \\
  \hline
  \hline
  \end{tabular}
  \caption{
     Superfluid parameters $\Delta$, $\mu$, and $m^*$ obtained by
     fitting DMC values of $\varepsilon(k)$ to Eq.\@
     1, along with DMC estimates of the
     condensate fraction $c$ and pair size $r_{\rm ex}$.
     \label{Stable:epsk_fit_dmc}}
\end{table}

\section{Finite-size tests}

The results reported in the main manuscript have been obtained by
simulating a 58-exciton system.
In this section we quantify the bias incurred by this choice of
system size.
We expect this bias to be small, since the superfluid parameters
are obtained via $\varepsilon(k)$, which is a difference of energy
differences, $\varepsilon(k) = E_A(N+1/2;k) - E_A(N) -
\frac 1 4 \left[ E_A(N+1) - E_A(N-1) \right]$,
potentially enabling substantial error cancellations in the
final results.

\begin{figure}[htb!]
  \includegraphics[width=0.48\textwidth]{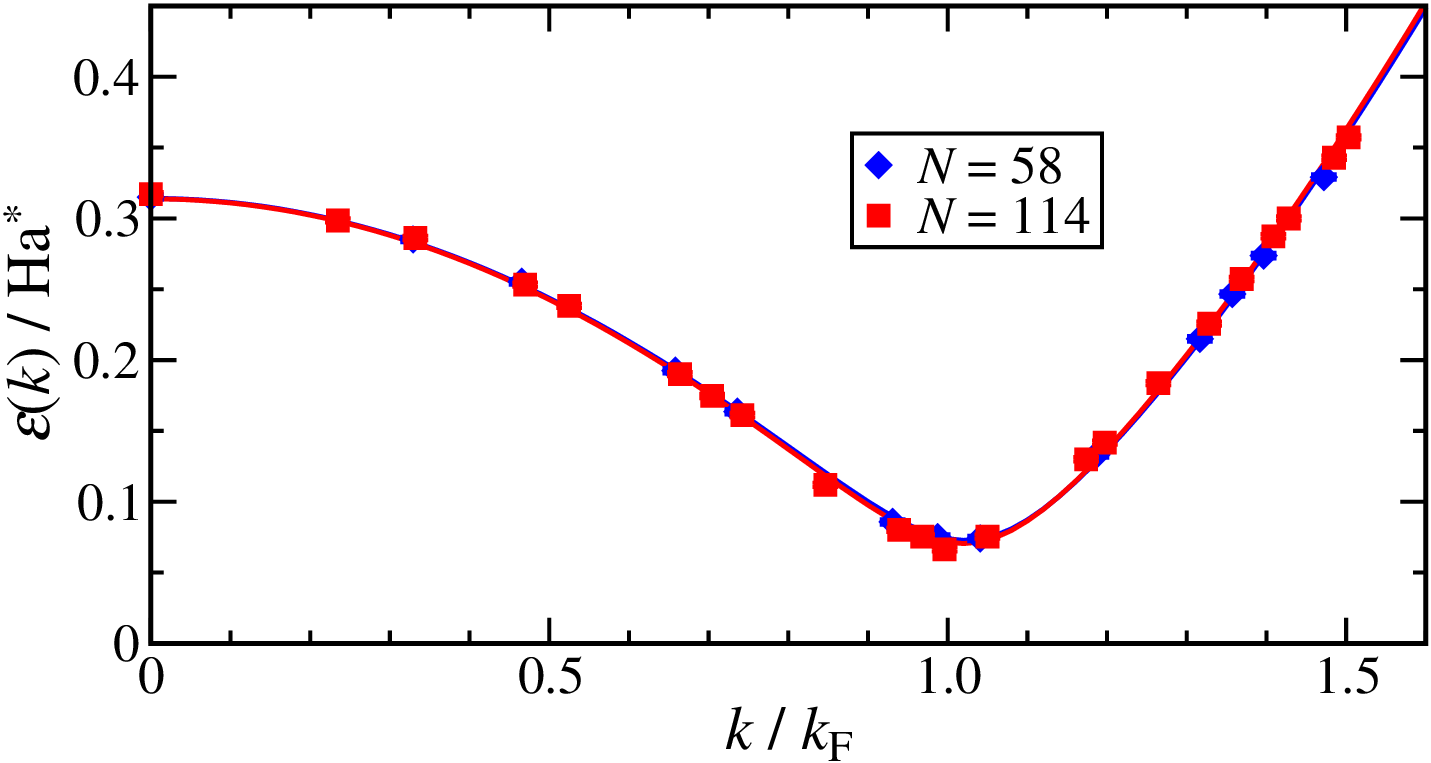}
  \caption{
    Excitation energies at $r_{\rm s}/a_0^*=2$ computed from VMC
    simulations of systems of $N=114$ electron-hole pairs (red
    squares) and $N=58$ electron-hole pairs (blue diamonds).
  }
  \label{Sfig:vmc114_epsk_rs2}
\end{figure}

We have performed VMC tests using a simulation cell containing
$N=114$ electron-hole pairs at selected densities of $r_{\rm s}/a_0^*
= 2$, $5$, and $10$ to quantify the finite-size error in our results
at $N=58$.
To illustrate these results we plot the $N=114$ VMC excitation
energies for $r_{\rm s}/a_0^* = 2$ as a function of $k$ in Fig.\@
\ref{Sfig:vmc114_epsk_rs2}.
The VMC values of $\varepsilon(k)$ at both system sizes appear to lie
on the same curve as a function of $k/k_{\rm F}$.

\begin{figure}[htb!]
  \includegraphics[width=0.48\textwidth]{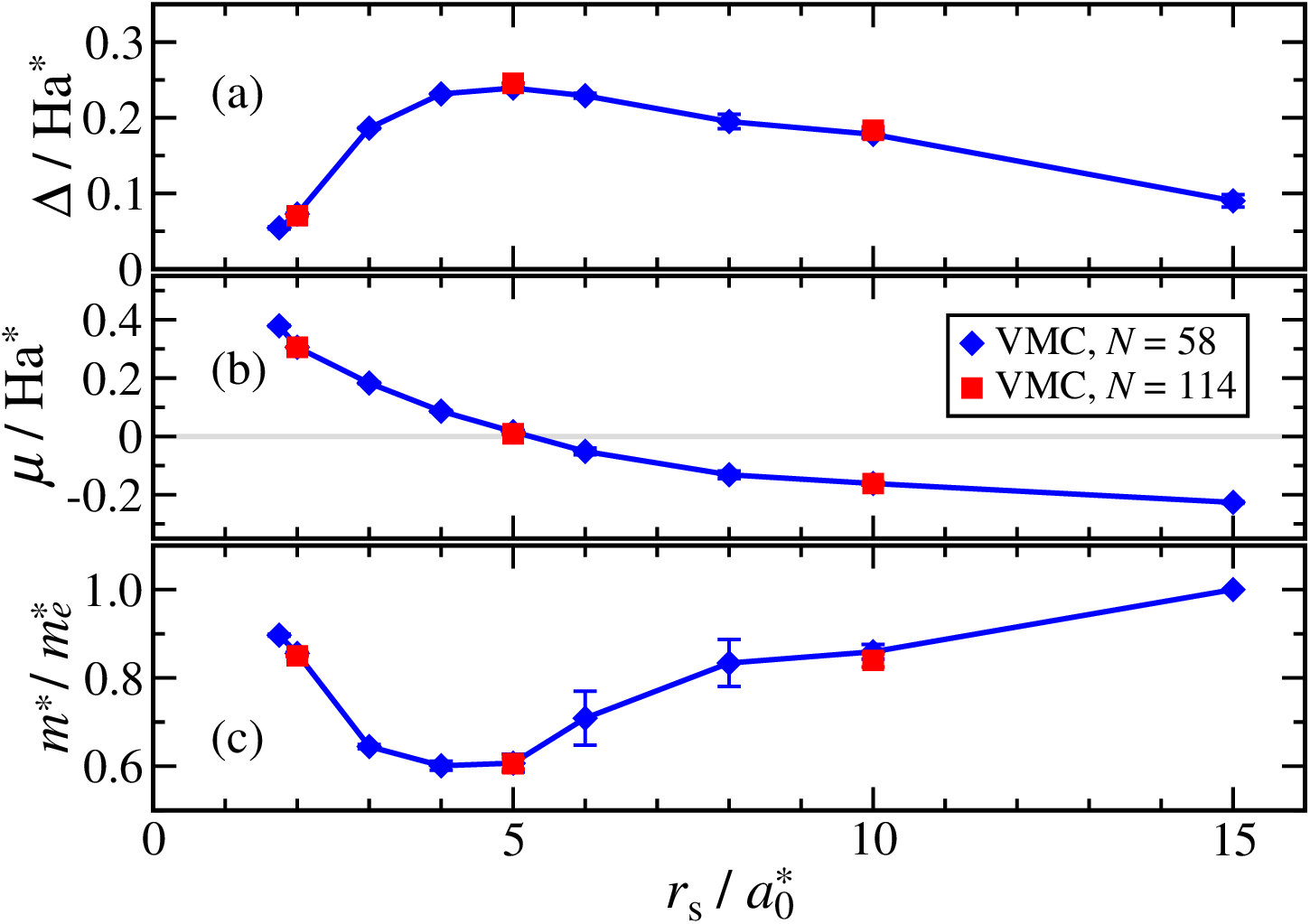}
  \caption{
    Superfluid parameters $\Delta$ (top panel), $\mu$ (middle
    panel), and $m^*$ (bottom panel) as a function of $r_{\rm s}$
    obtained from VMC simulations of systems of $N=114$
    electron-hole pairs (red squares) and $N=58$ electron-hole pairs
    (blue diamonds).
  }
  \label{Sfig:vmc114_params}
\end{figure}

The fit parameters obtained from the $N=114$ VMC data are tabulated
in Table \ref{Stable:epsk_fit_large} and plotted in Fig.\@
\ref{Sfig:vmc114_params}.
The superfluid parameters obtained for $N=114$ are nearly identical
to those at $N=58$, and we conclude that the finite-size errors of
our results are negligible.

~

\begin{table}[htb!]
  \centering
  \begin{tabular}{r@{.}lr@{.}lr@{.}lr@{.}lr@{.}lr@{.}l}
  \hline
  \hline
    \multicolumn{2}{c}{$r_{\rm s}/a_0^*$}
    & \multicolumn{2}{c}{$\Delta/{\rm Ha}^*$}
    & \multicolumn{2}{c}{$\mu/{\rm Ha}^*$}
    & \multicolumn{2}{c}{$m^*/m_e^*$}
    & \multicolumn{2}{c}{$c$}
    & \multicolumn{2}{c}{$r_{\rm ex}/a_0^*$} \\
  \hline
    $ 2$&$00$ & $0$&$0705(7)$  & $ 0$&$3058(7) $ & $0$&$850(2) $
              & $0$&$1576(24)$ & $1$&$159$ \\
    $ 5$&$00$ & $0$&$2455(3)$  & $ 0$&$0091(29)$ & $0$&$606(12)$
              & $0$&$6405(3) $ & $1$&$587$ \\
    $10$&$00$ & $0$&$1834(4)$  & $-0$&$1616(47)$ & $0$&$840(15)$
              & $0$&$8485(2) $ & $1$&$774$ \\
  \hline
  \hline
  \end{tabular}
  \caption{
     Superfluid parameters $\Delta$, $\mu$, and $m^*$ obtained by
     fitting VMC values of $\varepsilon(k)$ for a system of $N=114$
     electron-hole pairs to Eq.\@ 1, along with
     VMC estimates of the condensate fraction $c$ and pair size
     $r_{\rm ex}$.
     \label{Stable:epsk_fit_large}}
\end{table}

~

\FloatBarrier

\end{document}